\documentclass[prb,twocolumn]{revtex4}

\usepackage{amsmath,graphicx,amssymb,psfrag}

\newcommand{\NN}{\nonumber\\}
\newcommand{\dd}{{\rm d}}
\newcommand{\simD}{\stackrel{\text{\tiny D}}{\sim}}
\newcommand{\vnabla}{\vct{\nabla}}
\newcommand{\U}[1]{\mathrm{#1}}
\newcommand{\sub}[1]{_{\mbox{\scriptsize#1}}}
\newcommand{\vct}[1]{\boldsymbol#1}

\begin{document}
\title{
Derivation of magnetic Coulomb's law for 
thin, semi-infinite solenoids
}
\author{Masao Kitano}
\email{kitano@kuee.kyoto-u.ac.jp}
\affiliation{
Department of Electronic Science and
Engineering, Kyoto University,
Katsura, Kyoto 615-8510, Japan}
\affiliation{CREST, Japan Science and Technology Agency, Tokyo 103-0028, Japan}
\date{\today}

\begin{abstract}
It is shown that the magnetic force between thin, semi-infinite solenoids
obeys a Coulomb-type law, which corresponds to that
for magnetic monopoles placed at the end points of each solenoid.
We derive the magnetic Coulomb law from the basic principles of
electromagnetism, namely from the Maxwell equations and the 
Lorentz force.
\end{abstract}


\maketitle
\section{Introduction}
A permanent magnet is an ensemble of microscopic magnetic moments
which are oriented along the magnetization direction.
A magnetic moment can be modeled
as a dipole, i.e., as a slightly displaced pair of magnetic monopoles 
with opposite polarities.
It is analogous to the electric dipole.
Another way of modeling is to consider each magnetic moment 
as a circulating current loop.
In terms of far fields, the dipole model and 
the loop-current model give exactly
the same magnetic field.
The latter model is more natural because
there are no magnetic monopoles found so far and
microscopic magnetic moments are always associated with
kinetic rotations such as orbital motions or
spins of electrons.
It also provides correct symmetries with respect to
the time and space inversions.

Normally we are interested in macroscopic quantities,
which are obtained by coarse-graining of microscopic
fields and source distributions \cite{jackson}.
When we coarse-grain the oriented ensemble of 
microscopic magnetic dipoles in a bar magnet,
we have a magnetic north pole at one end and a south pole at the
other end as shown in Fig.~1(a).
Contributions of the magnetic charges inside are canceled out 
through the spatial
average.

On the other hand, when we coarse-grain the microscopic loop currents,
we have macroscopic current which circulates around the
bar along the surface as shown in Fig.~1(c).
By mimicking the macroscopic current distribution by
a coil, we have an electromagnet that is equivalent to the permanent magnet.
In this model, the poles, or the ends of bar magnet play no special roles.
Again the current model is much more reasonable
because of the similarity with the equivalent electromagnet.

\begin{figure}[b]
\psfrag{(a)}{(a)}
\psfrag{(b)}{(b)}
\psfrag{(c)}{(c)}
\psfrag{dipole}[c][c]{{\footnotesize magnetic dipole}}
\psfrag{loop}[c][c]{{\footnotesize current loop}}
\includegraphics[scale=0.6]{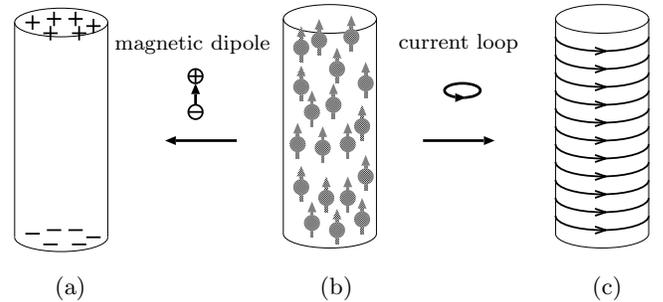}  
\caption{
A permanent magnet is an ensemble of microscopic magnetic
moments (b).
If we consider each magnetic moment as a magnetic dipole,
the corresponding macroscopic picture is two opposite magnetic poles at 
each end (a).
If we adopt the current loop model for magnetic moment,
the macroscopic picture consists of circulating current around
the side wall (c).
}
\end{figure}

The use of the notion of magnetic pole
should be avoided as far as possible 
\cite{warburton}
because of its absence in the framework of
electromagnetic theory or in the Maxwell equations.
Practically, however, magnetic poles are very
convenient to describe the forces between permanent
magnets or magnetized objects.
The poles are considered as an ensemble of monopoles
and the forces between poles are calculated with
the magnetic Coulomb law,
which is usually introduced just as an analog of
the electric Coulomb law
or as an empirical rule \cite{goldemberg}.
For logical consistency, we have to {\it derive}
the magnetic Coulomb law
from Maxwell's equations and the Lorentz force,
none of which contain the notion of magnetic monopoles.

The derivation of magnetic Coulomb's law was given
more than forty years ago
by Chen\cite{chen} and Nadeau\cite{nadeau}.
In this paper a more detailed analysis based
directly on the fundamental laws will be provided.
The field singularity for infinitesimal loop currents,
which plays crucial roles but was not mentioned 
in the previous works, will be treated rigorously.

\begin{figure*}
\psfrag{La}{$L\sub{a}$}
\psfrag{Lb}{$L\sub{b}$}
\psfrag{dl}{$\dd \vct l\sub{a}$}
\psfrag{kdl}{$\kappa\sub{a} \dd l\sub{a}$ loops}
\psfrag{m}{$\vct m$}
\psfrag{r1}{$\vct r_1$}
\psfrag{r2}{$\vct r_2$}
\psfrag{r3}{$\vct r_3$}
\psfrag{r4}{$\vct r_4$}
  \centering
\includegraphics[scale=0.8]{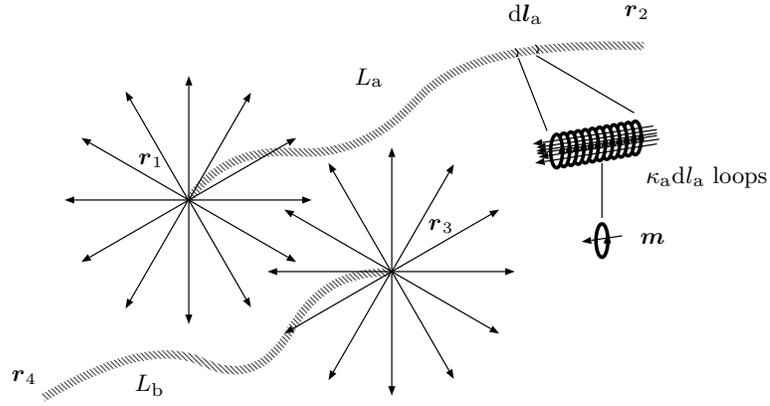}  
\caption{A pair of thin, semi-infinite solenoids,
$L\sub{a}$ and $L\sub{b}$}
\end{figure*}

\section{Current density for a thin solenoid}

For brevity, we introduce a scalar function $G_0$ and
a vector function $\vct G_1$ of position $\vct r=(x,y,z)$:
\begin{align}
  G_0(\vct r)=\frac{1}{4\pi|\vct r|},\quad
  \vct G_1(\vct r)=\frac{\vct r}{4\pi|\vct r|^3}
.
\label{eq10}
\end{align}
We note that ${\vnabla}G_0=-\vct G_1$
and ${\vnabla}\cdot\vct G_1=\delta^3(\vct r)$ hold,
where 
$\delta^3(\vct r)=\delta(x)\delta(y)\delta(z)$ is 
the three dimensional delta function.
With these, the scalar potential for a point charge
$q$ placed at the origin is
$\phi(\vct r)=(q/\varepsilon_0)G_0(\vct r)$,
and the force acting on a charge $q_1$ at
$\vct r_1$ from another charge $q_2$ at
$\vct r_2$ is
$\vct F_{1\leftarrow2}=(q_1 q_2/\varepsilon_0)\vct G_1(\vct r_1 - \vct r_2)$.
The Biot-Savard law can be expressed as
$\dd\vct H = \dd\vct C\times\vct G_1(\vct r)$,
where $\dd\vct C =I\dd\vct l$ is a current moment
(current $I$ times length $\dd\vct l$) located at the origin.

As shown in Fig.~2, a thin solenoid can be constructed as a stack of
tiny loop currents at least in principle \cite{helix}.
The current density for an infinitesimal loop current place at the
origin is
\begin{align}
  \vct J_{\vct m}(\vct r) = (-\vct m\times{\vnabla})\delta^3(\vct r)
\label{eq20}
\end{align}
where $\vct m$ is the
magnetic moment for the loop current and its unit
is  $(\U{A\,m^2})$ (See Appendix).
To form a solenoid we stack these tiny loop currents along a
curve $L\sub{a}$ with a constant line density
$\kappa\sub{a}$ (loops per unit length).
Each loop are aligned so that
the direction of tangent of curve and that of $\vct m$
coincide.

The current density distribution
for a line segment $\dd\vct l\sub{a}\,(\,\parallel\dd\vct m\,)$ is
\begin{align}
 \dd\vct J(\vct r) = (-C\sub{a}\dd\vct l\sub{a}\times{\vnabla})
   \delta^3(\vct r-\vct r\sub{a})
,
\label{eq30}
\end{align}
where
$C\sub{a}=\kappa\sub{a} m \simD \U{A\,m}$ is the
magnetic moment per unit length and characterizes
the strength of the solenoid \cite{equiv},
where $m=\vct m\cdot\dd\vct l\sub{a}/|\dd\vct l\sub{a}|$ is
the magnitude of $\vct m$ and
$\vct r\sub{a}$ represents the position of
the segment $\dd\vct l\sub{a}$.

\section{Field by a thin, semi-infinite solenoid}
Here we introduce an important formula.
For a constant vector $\vct{\alpha}$ and a vector field
$\vct{V}(\vct r)$, we have
\begin{align}
(\vct{\alpha}\times{\vnabla})\times\vct{V}
&-\vct{\alpha}\times({\vnabla}\times\vct{V})
\NN
&=(\vct{\alpha}\cdot{\vnabla})\vct{V}
-\vct{\alpha}({\vnabla}\cdot\vct{V})  
.
\label{eq35}
\end{align}

With the Biot-Savard law
and Eq.~(\ref{eq20}), we can
find the strength of magnetic field
generated by a magnetic moment
$\vct m$ placed at the origin as
\begin{align}
  \vct H_{\vct m}(\vct r) &= 
\int\dd v'
\vct J_{\vct m}(\vct r')\times
\vct G_1(\vct r - \vct r')
\NN
&=
 (- \vct m \times {\vnabla})\times\vct G_1(\vct r)
,
\label{eq38}
\end{align}
where $\dd v'$ is a volume element at $\vct r'$
and ${\vnabla}'=\partial/\partial\vct r'$.
Using Eq.~(\ref{eq35}), it can be rewritten as
\begin{align}
  \vct H_{\vct m}(\vct r) = 
-(\vct m \cdot {\vnabla})\vct G_1(\vct r)
+\vct m \delta^3(\vct r)
,
\label{eq40}
\end{align}
where the relations
${\vnabla}\cdot\vct G_1(\vct r)
=\delta^3(\vct r)$
and ${\vnabla}\times\vct{G}_1=0$ have been used.
As shown in Appendix, Eq.~(\ref{eq40}) can also be derived from
the Maxwell equations.

It is well known that the field for an electric dipole moment
$\vct p$ is 
$\vct D(\vct r)=-(\vct p\cdot {\vnabla})\vct G_1(\vct r)$,
which contains no delta-function terms unlike Eq.~(\ref{eq40}).
This means that 
in terms of near fields, the dipole and the current loop do not
yield the same field.

From Eq.~(\ref{eq40}),
the strength of the magnetic field
created by a line segment $\dd\vct l\sub{a}$ is
\begin{align}
  \dd\vct H(\vct r)= 
     (C\sub{a}\dd\vct l\sub{a}\cdot{\vnabla}\sub{a})
     \vct G_1(\vct r - \vct r\sub{a})
  + C\sub{a}\dd\vct{l}\sub{a}\delta^3(\vct r - \vct r\sub{a})
\label{eq60}
\end{align}
where ${\vnabla}\sub{a}=\partial/\partial\vct r\sub{a}$.
The integration along a curve $L\sub{a}$ yields
the magnetic field created by the solenoid:
\begin{align}
\vct H(\vct r)
&=\int_{L\sub{a}}\dd\vct H
= C\sub{a}\left[
  \vct G_1(\vct r - \vct r_1)
 -\vct G_1(\vct r - \vct r_2)
\right]
\NN
{}&+C\sub{a}\int_{L\sub{a}} \delta^3(\vct r - \vct r\sub{a})\dd\vct l\sub{a}
,
\label{eq70}
\end{align}
where
$\vct r_2$ and $\vct r_1$ are the start and the end points of $L\sub{a}$.
The second term of the right-hand side, which corresponds to
the magnetic flux confined in the solenoid,
vanishes outside.
For semi-infinite cases ($\vct r_2=\infty$), we have
\begin{align}
  \vct H(\vct r) = 
  C\sub{a}\vct G_1(\vct r - \vct r_1)
  +C\sub{a}\int_{L\sub{a}} \delta^3(\vct r - \vct r\sub{a})\dd\vct l\sub{a}
.
\label{eq80}
\end{align}
The first term of the right-hand side,
which represents the magnetic field outside of the solenoid,
is equivalent to the
field for a monopole $g\sub{a}=\mu_0 C\sub{a}$ 
located at $\vct r_1$:
\begin{align}
  \vct B(\vct r)=g\sub{a}\vct G_1(\vct r - \vct r_1)
=\frac{g\sub{a}}{4\pi}\frac{\vct r - \vct r_1}{|\vct r - \vct r_1|^3}
.
\label{eq90}
\end{align}
The dimension of $g\sub{a}$,
\begin{align}
g\sub{a} = \mu_0 C\sub{a}  \simD \frac{\U{H}}{\U{m}}\,\U{A\,m}=
\frac{\U{V\,s/A}}{\U{m}}\,\U{A\,m}=\U{V\,s}=\U{Wb}
\label{eq100}
\end{align}
correctly corresponds to that for the magnetic charge.
The magnetic flux $g\sub{a}$ confined along the solenoid
fans out isotropically from the end point $\vct r_1$.
As seen in Fig.~2, a thin, semi-infinite solenoid can be viewed
as a magnetic monopole located at the end.

\section{Magnetic force acting on a semi-infinite solenoid}
From Eq.~(\ref{eq20}),
we see that the Lorentz force acting on a tiny loop current
$\vct m$ placed at $\vct r$ in a magnetic
field $\vct{B}$ is
\begin{align}
\vct F_{\vct m} &=
\int\dd v' \vct J_{\vct m}(\vct r')
 \times\vct B(\vct r + \vct r')
\NN
&=(\vct m\times{\vnabla})\times\vct B(\vct r)
.
\label{eq106}
\end{align}
Using Eq.~(\ref{eq35}) and 
the conditions for magnetic field:
${\vnabla}\cdot\vct B=0$ (divergence-free)
and 
${\vnabla}\times(\mu_0^{-1}\vct B)=0$ (rotation-free),
the expression can be modified as
\begin{align}
\vct F_{\vct m} =
(\vct m\cdot{\vnabla})\vct B(\vct r)
,
\label{eq110}
\end{align}
which is suitable for line-integral.
Equation (\ref{eq106}) can also be modified as
$\vct F_{\vct m} ={\vnabla}(\vct m\cdot\vct  B)$
with the divergence-free condition only.
The rotation-free condition is satisfied only when
the right hand side of the Maxwell-Amp\`{e}re equation,
${\vnabla}\times\vct H=\vct J+\partial\vct D/\partial t$,
vanishes.

Thus the magnetic force acting on a line element $\dd\vct l\sub{b}$
at $\vct r\sub{b}$ is
\begin{align}
  \dd\vct F = 
 (C\sub{b}\dd\vct l\sub{b}\cdot{\vnabla}\sub{b})\vct B(\vct r\sub{b})
,
\label{eq120}
\end{align}
where $C\sub{b}=\kappa\sub{b}m$ and 
${\vnabla}\sub{b}=\partial/\partial \vct r\sub{b}$.
Integration along a curve $L\sub{b}$ yields
the total force acting on the solenoid;
\begin{align}
  \vct F &= \int_{L\sub{b}} \dd\vct F
=C\sub{b}\int_{L\sub{b}} (\dd \vct l\sub{b}\cdot{\vnabla}\sub{b})\vct B
\NN
&=C\sub{b}\left[\vct B(\vct r_3) - \vct B(\vct r_4)\right]
,
\label{eq130}
\end{align}
where
$\vct r_4$ and $\vct r_3$ are the
initial and end points of $L\sub{b}$,
respectively.
For semi-infinite cases [$\vct B(\vct r_4=\infty)=0$], we have
\begin{align}
  \vct F = C\sub{b}\vct B(\vct r_3)
= g\sub{b}\vct H(\vct r_3)
,
\label{eq140}
\end{align}
with $\vct H =\mu_0^{-1}\vct B$.
This is equal to
the force for a point magnetic charge
$g\sub{b}=\mu_0C\sub{b}$ placed at $\vct r_3$.

It is surprising that the
sum of the forces acting on each part of the solenoid
can be represented in terms only of
the magnetic field $\vct B(\vct r_3)$
at the end point $\vct r_3$.
This is because the partial force
is proportional to the (vector) gradient of
the magnetic field.

\section{Coulomb's law between two thin, semi-infinite solenoids}
Now we can calculate the force (\ref{eq140})
on solenoid b by the field (\ref{eq80}) generated by
solenoid a;
\begin{align}
  \vct F\sub{b$\leftarrow$a} &= C\sub{b}\vct B(\vct r_3)
= C\sub{b}\mu_0\vct H(\vct r_3)
\NN
&= \mu_0 C\sub{b}C\sub{a}
\vct G_1(\vct r_3 - \vct r_1)
.
\label{eq150}
\end{align}
This equation corresponds to
Coulomb's law for
magnetic charges $g\sub{a}$ at $\vct r_1$ and $g\sub{b}$
at $\vct r_3$:
\begin{align}
  \vct F_{3\leftarrow1} = \frac{g\sub{a}g\sub{b}}{\mu_0}
  \vct G_1(\vct r_3 - \vct r_1)
  =\frac{g\sub{a}g\sub{b}}{4\pi\mu_0}\frac{\vct r_3 - \vct r_1}
  {|\vct r_3 - \vct r_1|^3}
.\label{eq160}
\end{align}
These two equations are exactly the same
but the former is for the total (integrated) force between
currents of solenoids and the latter is 
for the force between magnetic point charges.

It is interesting that the force is
independent of the paths of either solenoids as long as
their end points and strengths are kept constant.
We should note that the derivation of the magnetic
Coulomb law is not straightforward.
The properties of static, source-free fields must be
utilized.

\section{Torque on a semi-infinite solenoid}
Now we are interested in the torque due to the integrated
magnetic force on a thin, semi-infinite solenoid.
The torque $\vct N_{\vct m}$ 
with respect to the origin exerted 
on a loop current $\vct m$ located at
$\vct r$ is \cite{nadeau}
\begin{align}
\vct N_{\vct m}
&=\int
\dd v'
(\vct r + \vct r')
\times
\left[
\vct J_{\vct m}(\vct r')
\times
\vct B(\vct r + \vct r')
\right]
\NN
&=
\vct r\times \vct F_{\vct m}
+\int\dd v'
\left[
\vct B(\vct r + \vct r')\times\vct J_{\vct m}(\vct r')
\right]
\times\vct r'
\NN
&= \vct N^{(0)}_{\vct m} + \vct N^{(1)}_{\vct m} 
,
\end{align}
where the first term corresponds to the torque due to
the total force $\vct F_{\vct m}$ with respect to the origin and
the second term corresponds to the torque with respect to the
center of the loop current.
Using Eq.~(\ref{eq110}), the former can be modified as
\begin{align}
\vct N^{(0)}_{\vct m} &=
\vct r\times\vct F_{\vct m}
=\vct r\times\vct (\vct m\cdot{\vnabla})\vct B
\NN
&=(\vct m\cdot{\vnabla})(\vct r\times\vct B)
-[(\vct m\cdot{\vnabla})\vct r]\times\vct B
\NN
&=(\vct m\cdot{\vnabla})(\vct r\times\vct B)
-\vct m\times\vct B
,
\end{align}
with $(\vct m\cdot{\vnabla})\vct r=\vct m$.
The latter can be simplified as
\begin{align}
\vct N^{(1)}_{\vct m} 
&=
\int\dd v'
\left[
\vct B(\vct r+\vct r')\times(-\vct m\times{\vnabla}')\delta^3(\vct r')
\right]
\times\vct r'
\NN
&=
\left.\vct r'\times[(\vct m\times{\vnabla}')\times\vct B(\vct r+\vct r')]
\right|_{\vct r'=0}
\NN
&\phantom{=ABC} {}+
\left.[\vct B(\vct r+\vct r')\times(\vct m\times{\vnabla}')]\times \vct r'
\right|_{\vct r'=0}
\NN
&=0 + \left.[(\vct B\cdot{\vnabla}')\vct m
       - (\vct B\cdot\vct m){\vnabla}']\times\vct r'\right|_{\vct r'=0}
\NN
&=\vct m\times \vct B
,
\end{align}
where we have utilized 
${\vnabla}'\times\vct r' = 0$
with $\vnabla'=\partial/\partial\vct r'$.
.
Finally, we have
\begin{align}
\vct N_{\vct m}=
\vct N^{(0)}_{\vct m}+
\vct N^{(1)}_{\vct m}=
(\vct m\cdot{\vnabla})
[\vct r\times\vct B(\vct r)]
.
\end{align}
The torque on a line element $\dd\vct l\sub{b}$
at $\vct r\sub{b}$ is
\begin{align}
\dd\vct N =
C\sub{b}(\dd\vct l\sub{b}\cdot{\vnabla}\sub{b})
[\vct r\sub{b}\times\vct B(\vct r\sub{b})]
,
\end{align}
and the integration along the semi-infinite solenoid
$L\sub{b}$ terminated at $\vct r_3$ yields
\begin{align}
\vct N &= \int_{L\sub{b}}\dd \vct N
=
C\sub{b}\int_{L\sub{b}}
(\dd\vct l\sub{b}\cdot{\vnabla}\sub{b})
[\vct r\sub{b}\times\vct B(\vct r\sub{b})]
\NN
&=C\sub{b}\vct r_3\times\vct B(\vct r_3)
=\vct r_3\times\vct F
,
\end{align}
where $\vct F$ is the force (\ref{eq140}) on the solenoid.
Surprisingly, the torque exerted on a thin, semi-infinite
solenoid coincides with that for a point magnetic charge
$g\sub{b}=\mu_0 C\sub{b}$ place at $\vct r_3$.

\section{Discussion}
In this paper we only deals with thin solenoids.
Solenoids with finite cross-section
can be represented
as a bundle of thin solenoids.
Similarly a permanent magnet can naturally be
modeled as a bundle of thin solenoids.
The degree of magnetization of magnets or magnetized objects
is characterized by the quantity called macroscopic 
{\it magnetization}, $M\simD \U{A/m}$,
which is defined as volume density of 
magnetic moments, $m\simD \U{A\,m^2}$.
The magnetization can also be represented as
area-density of thin solenoids.
We remember that a thin solenoid is characterized by the
current moment, $C\simD \U{A\,m}$.
Bundling $\nu$ solenoids per unit area ($\simD \U{/m^2}$), we can create
the magnetization $M=\nu C$.

In conclusion, we have shown that the force 
between thin, semi-infinite solenoids obeys the Coulomb law
that is for the equivalent magnetic point charges placed at each end
of solenoid.
We have also shown that the torque exerted on the solenoid in a magnetic
field coincides with that for a corresponding magnetic point charge.
It is convenient to introduce magnetic charges or poles
because the magnetic Coulomb law can easily be applied for forces
between them.
But we should remember that 
without the justification given in this paper 
it is just a rough and ready method.

\begin{acknowledgments}
This work is supported by the 21st Century COE program No.~14213201.
\ \\
\end{acknowledgments}

\appendix*
\section{Current density and field of a tiny current loop}
The current density (\ref{eq20}) for a tiny current loop
can be derived as follows.
We consider a parallelogram
defined by a pair of small vectors, $\vct a$ and $\vct b$.
The center is located at the origin and
current $I$ is circulating along the edge.
The current distribution on the four segments can be approximately represented
as
\begin{align}
\vct J_{I(\vct a\times\vct b)}(\vct r)
=& I\left[
\vct a\delta^3(-\vct b/2)
+\vct b\delta^3(\vct a/2)
\right.
\NN
&
\left.
{}-\vct a\delta^3(\vct b/2)
-\vct b\delta^3(-\vct a/2)
\right]  
\end{align}
In the limit of
$|\vct a|,|\vct b|\rightarrow0$
with $\vct m=I(\vct a\times\vct b)$ being kept
constant,
it approaches
\begin{align}
\vct J_{\vct m}(\vct r)
&=
I\left[
-\vct a(\vct b\cdot{\vnabla})\delta^3(\vct r)
+\vct b(\vct a\cdot{\vnabla})\delta^3(\vct r)
\right]
\NN
&=
[-I(\vct a\times\vct b)\times{\vnabla}]\delta^3(\vct r)
\NN
&=
(-\vct m\times{\vnabla})\delta^3(\vct r)
,
\label{eq520}
\end{align}
and Eq.~(\ref{eq20}) is obtained.
The derivative of the delta function $\vnabla\delta^3(\vct r)$ serves as 
a differential operator when it is integrated together with
other functions as in 1D cases:
$\int \dd x f(x) ({\dd}/{\dd x})\delta(x)=-({\dd f}/{\dd x})(0)$
.
For example, the magnetic force on $\vct m$ can be
calculated as follows;
\begin{align}
  \vct F_{\vct m}
&=\int \dd v (-\vct m\times\vnabla)\delta^3(\vct r)\times\vct B(\vct r)
\NN
&=\int \dd v
\left[
-\vnabla\delta^3(\vct m\cdot\vct B)+\vct m(\vnabla\delta^3\cdot\vct B)
\right]
\NN
&=\left[\vnabla(\vct m\cdot\vct B)-\vct m(\vnabla\cdot\vct B)
  \right]_{\vct r=0}
\NN
&=\left[(\vct m\times\vnabla)\times\vct B\right](0)
.
\end{align}
Next we derive Eq.~(\ref{eq40})
from Maxwell's equations.
The Amp\`{e}re law for an infinitesimal loop current,
${\vnabla}\times\vct H=\vct J_{\vct m}$ can be
written as
\begin{align}
  {\vnabla}\times\left(\vct H(\vct r) -\vct m\delta^3(\vct r)\right)=0
.
\label{eq540}
\end{align}
We know that a rotation-free field can be represented as a gradient of
some scalar field $\phi\sub{m}(\vct r)$ as
\begin{align}
 \vct H(\vct r) -\vct m\delta^3(\vct r)=-{\vnabla}\phi\sub{m}(\vct r)
.
\label{eq560}
\end{align}
Taking the divergence of each side, we have
\begin{align}
   ({\vnabla}\cdot\vct m)\delta^3(\vct r)=\nabla^2\phi\sub{m}(\vct r)
,
\label{eq580}
\end{align}
where ${\vnabla}\cdot(\mu_0\vct H)=0$ has been used.
Comparing it with
\begin{align}
  \nabla^2({\vnabla}\cdot\vct m)G_0(\vct r)
   =-({\vnabla}\cdot\vct m)\delta^3(\vct r)
,
\label{eq600}
\end{align}
which is obtained from
$\nabla^2G_0(\vct r) = -\delta^3(\vct r)$,
we find the solution of (\ref{eq580}) to be
\begin{align}
\phi\sub{m}(\vct r)=-({\vnabla}\cdot\vct m)G_0(\vct r)
.
\label{eq620}
\end{align}
With Eq.~(\ref{eq560}) and ${\vnabla}G_0=-\vct G_1$ 
we have Eq.~(\ref{eq40}):
\begin{align}
\vct H(\vct r)=
-(\vct m\cdot{\vnabla})\vct G_1(\vct r)+\vct m\delta^3(\vct r)
,
\label{eq640}
\end{align}
which can be confirmed to satisfy the Maxwell equations.


\begin{thebibliography}{9}

\bibitem{jackson}
J. D. Jackson, {\it Classical Electrodynamics}
(Addison Wesley and Sons, New York, 1998),
 3rd ed. 
\bibitem{warburton}
F. W. Warburton, ``The magnetic pole, A useless concept,''
Am. Phys. Teacher {\bf 2}, 1 (1934).
\bibitem{chen}
H. S. C. Chen, ``Note on the magnetic pole,''
Am. J. Phys. {\bf 33}, 563 (1965).
\bibitem{nadeau}
G. Nadeau: ``Comment on Chen's note on the
magnetic pole,'' Am. J. Phys. {\bf 34}, 60 (1966).
\bibitem{goldemberg}
J. Goldemberg, ``An experimental verification of the Coulomb law
for magnetic poles,'' Am. J. Phys. {\bf 20}, 591--592 (1952).
\bibitem{helix}
In practice, a solenoid is made as a helix
of a conducting wire not as a stack of closed
wire loops.
The difference in current distribution,
which can be represented as an additional current
along the solenoid, can be
made arbitrarily small by reducing the
wire current and increasing the number of
winding correspondingly.
\bibitem{equiv}
The relation $A\simD B$ means $A$ and $B$ are dimensionally equivalent
and is normally written as $[A]=[B]$.
\end{thebibliography}
\end{document}